\documentclass[journal]{IEEEtran}
\usepackage{amsmath,amsthm,amsfonts,yhmath}
\usepackage{amssymb}
\usepackage[normalem]{ulem}

\usepackage{tikz}
\usepackage{url}
\usepackage{cuted}
\usepackage[nolist]{acronym}
\usepackage{todonotes}

\newcommand{\rev}[1]{\textcolor{black}{#1}} 

\usetikzlibrary{%
	mindmap,
	trees,
	graphs,
  shapes.misc,
  shapes.arrows,%
  chains,%
  matrix,%
  positioning,
  scopes,%
  decorations.pathreplacing,
  shadows%
}

\newif\ifgreek
\def\testgreek#1{
  \ifx#1\alpha\greektrue\else
  \ifx#1\beta\greektrue\else
  \ifx#1\gamma\greektrue\else\ifx#1\Gamma\greektrue\else
  \ifx#1\delta\greektrue\else\ifx#1\Delta\greektrue\else
  \ifx#1\epsilon\greektrue\else
  \ifx#1\zeta\greektrue\else
  \ifx#1\eta\greektrue\else
  \ifx#1\theta\greektrue\else\ifx#1\Theta\greektrue\else
  \ifx#1\iota\greektrue\else
  \ifx#1\kappa\greektrue\else
  \ifx#1\lambda\greektrue\else\ifx#1\Lambda\greektrue\else
  \ifx#1\mu\greektrue\else
  \ifx#1\nu\greektrue\else
  \ifx#1\xi\greektrue\else\ifx#1\Xi\greektrue\else
  \ifx#1\pi\greektrue\else\ifx#1\Pi\greektrue\else
  \ifx#1\rho\greektrue\else
  \ifx#1\sigma\greektrue\else\ifx#1\Sigma\greektrue\else
  \ifx#1\tau\greektrue\else
  \ifx#1\upsilon\greektrue\else\ifx#1\Upsilon\greektrue\else
  \ifx#1\phi\greektrue\else\ifx#1\Phi\greektrue\else
  \ifx#1\chi\greektrue\else
  \ifx#1\psi\greektrue\else\ifx#1\Psi\greektrue\else
  \ifx#1\omega\greektrue\else\ifx#1\Omega\greektrue\else
  \ifx#1\varepsilon\greektrue\else
  \ifx#1\vartheta\greektrue\else
  \ifx#1\varrho\greektrue\else
  \ifx#1\varsigma\greektrue\else
  \ifx#1\varphi\greektrue\else
     \greekfalse
  \fi\fi\fi\fi\fi\fi\fi\fi\fi\fi
  \fi\fi\fi\fi\fi\fi\fi\fi\fi\fi
  \fi\fi\fi\fi\fi\fi\fi\fi\fi\fi
  \fi\fi\fi\fi\fi\fi\fi\fi\fi}

\renewcommand{\vec}[1]{{\boldsymbol#1}}  
\newcommand{\mat}[1]{{\testgreek#1\ifgreek\boldsymbol#1\else
                      \mathbf#1\fi}} 

\newcommand{\ie}{\textit{i.e.}\/, }  
\newcommand{\eg}{\textit{e.g.}\/, }

\providecommand*{\mrm}[1]{\mathrm{#1}}

\DeclareMathAccent{\ring}{\mathalpha}{operators}{"17}

\providecommand*{\unit}[1]{\ensuremath{\mrm{\,#1}}}  

\renewcommand{\Re}{\operatorname{Re}}	
\providecommand*{\diff}{\operatorname{d}\!}  

\newlength{\temp}
\newlength{\tempa}

\newcommand{\Um}{\mat{U}}

\newcommand{\Jm}{\mat{I}}
\newcommand{\Tm}{\mat{T}}

\newcommand{\Pm}{\mat{P}}
\newcommand{\Pmt}{\widetilde{\mat{P}}}

\newcommand{\Bm}{\mat{B}}

\newcommand{\Hm}{\mat{H}}
\newcommand{\Om}{\mat{0}}
\newcommand{\Vm}{\mat{V}}
\newcommand{\Zm}{\mat{Z}}

\newcommand{\Rm}{\mat{R}}
\newcommand{\Sm}{\mat{S}}

\newcommand{\Sigmam}{\mat{\Sigma}}

\newcommand{\Psim}{\mat{\Psi}}

\newcommand{\Rrm}{\mat{R}_{\mrm{r}}}
\newcommand{\Rml}{\mat{R}_{\mrm{\Omega}}}

\newcommand{\xm}{\mat{x}}
\newcommand{\ym}{\mat{y}}
\newcommand{\nm}{\mat{n}}

\newcommand{\Upsilonm}{\mat{\Upsilon}}
\newcommand{\Rmn}{\mat{R}_{\nu}}

\newcommand{\Pd}{P_{\mrm{d}}}
\newcommand{\Pl}{P_{\mrm{\Omega}}}

\newcommand{\herm}{\text{H}}

\newcommand{\medel}[1]{\mathcal{E}\left\{ #1\right\}}

\newcommand{\Tr}{\mathop{\mrm{Tr}}\nolimits}

\newcommand{\eig}{\mathop{\mrm{eig}}\nolimits}
\newcommand{\svd}{\mathop{\mrm{svd}}\nolimits}

\newcommand{\maximize}{\mrm{maximize}}

\newcommand{\subto}{\mrm{subject\ to}}

\newcommand{\Id}{\mat{1}}

\newcommand{\reg}{\varOmega}

\newcommand{\Hmt}{\widetilde{\mat{H}}}

\providecommand{\V}[1]{\boldsymbol{#1}}
\providecommand{\basfcn}{\V{\psi}}
\newcommand{\sigman}{\sigma_n}

\newcommand{\Rs}{R_{\mrm{s}}}
\newcommand{\rhon}{\varrho_{n}}

\tikzset{%
  wireless/.pic={
      \draw [->] (0,0) -| (.5,#1);
    \foreach \r in {.1,.2,.3}
      \draw (.6,#1) ++ (60:\r) arc (60:-60:\r);
  },
  wirelessr/.pic={
      \draw [->] (0.0,0) -| (.5,#1);
    \foreach \r in {.1,.2,.3}
      \draw (.99,#1) ++ (240:\r) arc (240:120:\r);
  },
  vdots/.pic={
    \foreach \i in {-.1,0,.1}
      \fill (.25,\i) circle [radius=.75pt]; 
  },
  block/.style={
    shape=rectangle,
    minimum width=2cm,
    minimum height=1cm,
    draw
  },
  Tx/.style 2 args={
    block,
    node contents=Tx,
    append after command={
      \pgfextra{\pgfnodealias{@}{\tikzlastnode}}
      (@.north #1) [yshift=-.125cm] pic [#2] {wireless=.5}
      (@.#1)                        pic [#2] {vdots}
      (@.south #1) [yshift= .125cm] pic [#2] {wireless=.5}
    }
  },
  Rx/.style 2 args={
    block,
    node contents=Rx,
    append after command={
      \pgfextra{\pgfnodealias{@}{\tikzlastnode}}
      (@.north #1) [yshift=-.125cm] pic [#2] {wirelessr=.5}
      (@.#1)                        pic [#2] {vdots}
      (@.south #1) [yshift= .125cm] pic [#2] {wirelessr=.5}
    }
  },  
  MIMO Rx east/.style={Rx={east}{xscale=1}},
  MIMO Tx west/.style={Tx={west}{xscale=-1}},
}

\newacro{MoM}[MoM]{method of moments}
\newacro{PEC}[PEC]{perfect electric conductor}
\newacro{EFIE}[EFIE]{electric field integral equation}
\newacro{MFIE}[MFIE]{magnetic field integral equation}
\newacro{FBW}[FBW]{fractional bandwidth}
\newacro{MIMO}[MIMO]{multiple-input-multiple-output}
\newacro{SNR}[SNR]{signal-to-noise ratio}
\newacro{SVD}[SVD]{singular value decomposition}

\title{Physical bounds and radiation modes for MIMO antennas}

\author{Casimir Ehrenborg,~\IEEEmembership{Student member,~IEEE}, and Mats Gustafsson,~\IEEEmembership{Senior member,~IEEE}
\thanks{Manuscript received \today. This work was supported by the Swedish foundation for strategic research (SSF) under the program applied mathematics and the project Complex analysis and convex optimization for electromagnetic design.}%
\thanks{C. Ehrenborg and M. Gustafsson are with the Department of Electrical and Information Technology, Lund University, Box 118, SE-221 00 Lund, Sweden. (Email: \{casimir.ehrenborg, mats.gustafsson\}@eit.lth.se).}
}

\begin{document}

\maketitle

\begin{abstract}
Modern antenna design for communication systems revolves around two extremes: devices, where only a small region is dedicated to antenna design, and base stations, where design space is not shared with other components. Both imply different restrictions on what performance is realizable. In this paper properties of both ends of the spectrum in terms of MIMO performance is investigated. For \rev{electrically} small antennas the size restriction dominates the performance parameters. The regions dedicated to antenna design induce currents on the rest of the device. Here a method for studying fundamental bound on spectral efficiency of such configurations is presented. \rev{This bound is also studied for $N$-degree MIMO systems.} For \rev{electrically} large structures the number of degrees of freedom available per unit area is investigated for different shapes. Both of these are achieved by formulating a convex optimization problem for maximum spectral efficiency in the current density on the antenna. A \rev{computationally efficient }
solution for this problem is formulated and investigated in relation to constraining parameters, such as size and efficiency.
\end{abstract}


\section{Introduction}\label{sec:intro}
In modern communication technology the use of several antennas organized in \ac{MIMO} systems have become ubiquitous. This enables much greater bit rate (capacity) to be sent through the link between device and base station~\cite{Molisch2011,Paulraj+etal2003}. 
Within smaller devices, such as hand held electronics, the space allotted for antenna design is extremely limited. Here, both space and power efficiency needs to be utilized as effectively as possible. It is therefore of interest to investigate fundamental bounds on performance of \ac{MIMO} systems, both in terms of size and efficiency. Previously, bounds on capacity has been investigated for spherical geometries~\cite{AlayonGlazunov+etal2011,Gustafsson+Nordebo2007a}, and through information theoretical approaches~\cite{Migliore2008,Franceschetti+etal2009,Taluja+Hughes2012,Kundu2016,Migliore2006a}. Characteristic modes have been utilized to design antennas for maximum capacity and diversity~\cite{Li+etal2014,Chen+Wang2015}. However, this does not solve the issue of predicting the optimal performance available through antenna design in an arbitrary geometry. A method for calculating optimal spectral efficiency available in an arbitrary volume was presented in~\cite{Ehrenborg+Gustafsson2018}. However, in most applications it is effective to use only a small region of the device volume to excite currents over the entire device~\cite{Hannula+etal2018}. Therefore bounding the performance of such sub-regions and finding their optimal placement is of interest.

Current optimization has been utilized to construct fundamental bounds on many different antenna parameters previously, such as, Q, directivity, and efficiency~\cite{Ehrenborg+Gustafsson2018,Gustafsson+Nordebo2013,Gustafsson+etal2015b,Gustafsson+etal2016a,Jelinek+Capek2017}. By controlling the current density in the full design space of the antenna an optimal solution can be reached for that configuration. The power of this method comes from the ability to formulate these optimization problems as convex optimization problems. This means that all local minima of the problem are also global minima~\cite{Boyd+Vandenberghe2004}. Therefore there is no risk of getting caught in local minima and the optimality of the solution can be guaranteed~\cite{Boyd+Vandenberghe2004}. This method works very well for single feed, single resonance antennas where their performance, in \eg Q-factor, can be calculated as simple quadratic forms~\cite{Gustafsson+etal2016a}. Such expressions can be evaluated very efficiently using eigenvalue expressions that enables their optimization~\cite{Capek+etal2017b}. The performance of \ac{MIMO} antennas on the other hand is usually quantified in terms of capacity, which is not calculated through a quadratic form in the current density, but through a log-determinant of the covariance of the current density. Such an optimization problem is a semi-definite optimization program in the covariance of the current distribution, which has, in general, one more order of unknowns~\cite{Boyd+Vandenberghe2004}. To solve such optimization problems the number of unknowns need to be reduced~\cite{Ehrenborg+Gustafsson2018}.

In this paper the method presented in~\cite{Ehrenborg+Gustafsson2018} is reformulated in order to find a \rev{computationally efficient}
expression \rev{for}
its solution. A convex optimization problem to maximize the spectral efficiency of an arbitrary transmitter antenna in an ideal channel is formulated in the current density on the antenna. The problem is restricted by the allowed ohmic loss in the structure and normalized by the radiated or dissipated power. This problem is solved by utilizing the good properties of the matrices calculating radiated power and ohmic losses. The solution to this problem is \rev{mainly} dependent on the radiation modes, which are modes that maximize the fraction between the radiated power and the power dissipated in ohmic losses~\cite{Schab2016}. These modes are dependent on the geometrical structure of the object they are induced over, therefore the designer has control over how well they can be induced. The relative strength of these modes\rev{, given by their normalized eigenvalues,} is investigated for a sphere, a cylinder, a disc, and a plate and its sub-regions. This information is used to analyze how many sub-regions are required to \rev{feed a $N$-degree \ac{MIMO} system and to} fully utilize the potential of a plate design region. The optimal spectral efficiency of the sub-region configurations is compared to \rev{an ideal, lossless, system with equal power allocation and} optimizing the currents over the plate and sphere.  \rev{Electrically} large structures and their number of viable radiation modes are also investigated.

The paper is organized in the following way. In Sec.~\ref{sec:MIMO} the \ac{MIMO} system studied in this paper is introduced and its convex optimization problem is stated. The dual of that problem is formulated and solved in Sec.~\ref{sec:dual}, creating \rev{an upper} bound for the original problem. \rev{This is followed by an analysis of how the radiation modes affect the optimal solution in Sec.~\ref{sec:radmodes}.} In Sec.~\ref{sec:Results} the results of the optimization problem is illustrated in several examples split into the following sub-sections: Sec.~\ref{sec:Mode_strength} where the radiation modes of different shapes are studied, Sec.~\ref{sec:Plate_sub-regions} where the radiation modes of a plate are excited by sub-regions, Sec.~\ref{sec:optimal_cap} where the optimal spectral efficiency of several structures is investigated, and Sec.~\ref{sec:mode_availability} where the mode availability of \rev{electrically} large structures is calculated. The paper is concluded in Sec.~\ref{sec:conclusions}.

\section{MIMO}\label{sec:MIMO}
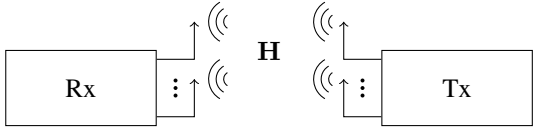
\begin{figure}%
\centering
\begin{tikzpicture}
\node at (0,0) [MIMO Rx east];
\node at (2.5,0.5) {$\mat{H}$};
\node at (5,0) [MIMO Tx west];
\end{tikzpicture}
\caption{Schematic illustration of a MIMO setup. Rx is the set of receiving antennas, Tx is the set of transmitting antennas, and $\Hm$ is the channel matrix describing the propagation between Tx and Rx.}%
\label{fig:MIMOschem}%
\end{figure}
\begin{figure}%
\centering
\includegraphics[width=0.7\columnwidth]{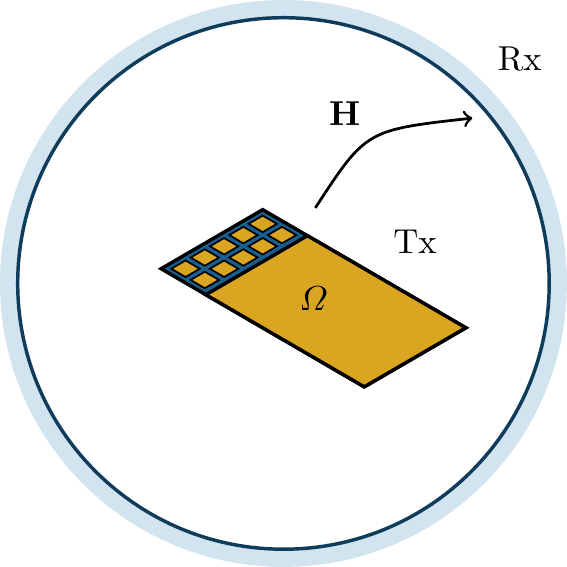}%
\caption{Schematic illustration of the idealized MIMO system where the receiving antennas are the spherical modes in the far-field. The transmitting antennas are illustrated as a plate with antenna design regions marked as small patches.}%
\label{fig:Id_channel}%
\end{figure}

The received signals in a \ac{MIMO} system is described by the expression,
\begin{equation}
\ym = \Hm\xm + \nm,
\label{eq:MIMO}
\end{equation}
where $\ym$ is an \rev{$M\times1$} vector containing the received signals, $\xm$ is an \rev{$N\times1$} vector containing the input signals, $\Hm$ is an \rev{$M\times N$} matrix describing the propagation channel, and $\nm$ \rev{is an $M\times1$ vector containing} the noise perturbing the system. In communication theory what is typically optimized is the power distribution in $\xm$ in order to send the maximum number of bits through a channel. However, in that configuration the antennas in the system are assumed fixed~\cite{Paulraj+etal2003}, see Fig.~\ref{fig:MIMOschem}. Here, we are interested in how much performance can be attained from optimizing those antennas for the specific application of transmitting the highest capacity. In order to calculate that performance, the problem must be reformulated slightly. First, the system in~\eqref{eq:MIMO} concerns two sets of antennas. However, in most cases we are not designing both the antennas in the system simultaneously, such as the base station and the mobile phone. Therefore, we reformulate the problem to optimizing a single device with regards to a general situation. In order to create such conditions, one set of antennas is idealized; here we choose the receiving antennas. Consider a receiving antenna completely circumscribing the transmitting antenna. This is similar to massive \ac{MIMO} and the intelligent surfaces discussed in~\cite{Hu+etal2017}, covering all surfaces of a room. In such a configuration all radiated energy would be absorbed by the receivers on the walls. This can be characterized by using the spherical modes in the far-field as the receiving antennas~\cite{AlayonGlazunov+etal2011,Gustafsson+Nordebo2007a,Ehrenborg+Gustafsson2018}, which will be used in this paper, see Fig.~\ref{fig:Id_channel}.

The second issue lies in the method of calculating optimal performance for the antennas. Since we do not know what shape an optimal antenna design would take, we want to incorporate every possible antenna in our solution space. To accomplish this we optimize the currents in the design space of the antenna, as these have the ability to represent every possible antenna within it. Normally a \ac{MIMO} system is optimized through controlling the input signals to the antennas. These input signals generate the currents on the antenna \rev{region $\varOmega$, see Fig.~\ref{fig:Id_channel}}, that are connected by a fixed mapping \rev{, $\Jm=\Tm\xm$}. However, since we are controlling the currents directly we simply utilize this map in order to calculate the performance quantities, such as capacity, from the currents~\cite{Ehrenborg+Gustafsson2018}. \rev{For brevity and ease of notation we do not write out this map in our expressions.} For example, instead of optimizing over the covariance of the input signals, we optimize over the covariance of the currents,
\begin{equation}
\Pm = \frac{1}{2}\medel{\Jm\Jm^{\herm}} .
\label{eq:current_covar}
\end{equation}
Essentially, each current element becomes an input signal to the system, whereas for actual antennas the number of inputs is restricted by the number of antenna feeds. 

The antenna region is modeled with a \ac{MoM} code, where each basis function corresponds to an element in $\Jm$. We call the matrix mapping the current distribution on the plate to spherical modes $\Sm$~\cite{Tayli+etal2018}, see App.~\ref{app:Smatrix}, which gives a new formulation of~\eqref{eq:MIMO},
\begin{equation}
\ym=\Sm\Jm + \nm .
\label{eq:sphMIMO}
\end{equation}
In this formulation the received signals $\ym$ are the radiated spherical modes, see Fig.~\ref{fig:Id_channel}. The average transmitted power of the system is calculated as,
\begin{multline}
P_{\mrm{r}} = \frac{1}{2}\medel{\Jm^{\herm}\Rrm\Jm} = 
\frac{1}{2}\Tr\medel{\Rrm\Jm\Jm^{\herm}} = \Tr(\Rrm\Pm) ,
\label{eq:average_power}
\end{multline}
\rev{where $\Rrm$ is the radiation matrix~\cite{Gustafsson+etal2016a}, and} we have utilized the cyclic properties of the trace. The capacity of this channel\rev{, when we have perfect channel state information,} expressed as spectral efficiency, is calculated as~\cite{Paulraj+etal2003}
\begin{equation}
  C = \max_{\Tr(\Rrm\Pm)=P_{\mrm{r}}}
  \log_2\det\left(\Id+\frac{1}{N_0}
      \Sm
      \Pm                                        
      \Sm^{\herm}
    \right),
\label{eq:capacity}
\end{equation} 
where $\Id$ is the identity matrix, and $N_0$ is the \rev{noise power, which can also be expressed as the noise spectral density over a $1\unit{Hz}$ frequency flat channel~\cite{Paulraj+etal2003}.}  The noise is modeled as white complex Gaussian noise. Maximum capacity and optimal energy allocation for~\eqref{eq:capacity} can be calculated by the water filling algorithm~\cite{Paulraj+etal2003}. However, it can also be written as an optimization problem solved by semidefinite programming,
\begin{equation}
\begin{aligned}
	& \maximize && \log_2\det\left(\Id+\gamma\Sm\Pm\Sm^{\herm}\right) &&&\\
	& \subto && \Tr(\Rrm\Pm) = 1 &&& \\
  & && \ \Pm \succeq 0 ,&&&
\end{aligned}
\label{eq:convexMIMO}	
\end{equation}
where the radiated power is normalized to one, and $\gamma = P_{\mrm{r}}/N_0$ is the total \ac{SNR}. If the \ac{SNR} is scaled with the number of included channels, in this case mesh discritization and number of spherical modes, the spectral efficiency is unbounded~\cite{Paulraj+etal2003,Ehrenborg+Gustafsson2018}. However, if the \ac{SNR} is fixed~\rev{\eqref{eq:convexMIMO}} converges \rev{to $\gamma/\log(2)$}
as \rev{$M,N\rightarrow\infty$, when the} mesh discretization and number of spherical modes \rev{are} increased. This problem formulation has the advantage that additional constraints can be added in order to gain a more realistic solution. Here, we can, for example, limit the losses in the structure. Loss is modeled as a uniform impedance sheet with surface resistance $\Rs$ in $\Omega/\square$ and calculated as,
\begin{equation}
  \Pl = \frac{1}{2}\medel{\Jm^{\herm}\Rml\Jm}
  =\Tr(\Rml\Pm) , 
\end{equation}
where $\Rml=\Rs\Psim$ is the loss matrix of the antenna~\cite{Gustafsson+etal2016a}, and $\Psim$ is the Gramian matrix of the \ac{MoM} basis functions on the antenna. Adding this constraint to~\eqref{eq:convexMIMO} gives the formulation
\begin{equation}
\begin{aligned}
	& \maximize && \log_2\det(\Id+\gamma\Sm\Pm\Sm^{\herm})\\
	& \subto && \Tr(\Rml\Pm) \leq \rev{\eta^{-1}-1} \\
	& && \Tr(\Rrm\Pm) = 1 \\ 
	& && \ \Pm \succeq \Om ,
\end{aligned}
\label{eq:convex_MIMOr}
\end{equation}
\rev{where $\eta$ is the radiation efficiency of the antenna.}
This problem is bounded and converges as the mesh discretization is refined, and the included number of spherical modes is increased~\cite{Ehrenborg+Gustafsson2018}. However, it contains many unknowns and is cumbersome to solve numerically. Because it is a semi-definite programming problem the number of unknowns scale as the square of the number of mesh cells, necessitating model order reductions or other 
\rev{numerical procedures} to run the optimization~\cite{Ehrenborg+Gustafsson2018}.

\rev{The optimization problem~\eqref{eq:convex_MIMOr} is normalized to radiated power, however, when losses are included it is sometimes intuitive to normalize dissipated power $P_{\mrm{d}} = P_{\mrm{r}}+P_{\Omega}$ instead. The problem is then written as,}
\begin{equation}
\begin{aligned}
	&\rev{ \maximize }&&\rev{ \log_2\det(\Id+\gamma\Sm\Pm\Sm^{\herm})}\\
	&\rev{ \subto }&&\rev{ \Tr(\Rml\Pm) \leq 1-\eta }\\
	& &&\rev{ \Tr((\Rrm+\Rml)\Pm) = 1 }\\ 
	& &&\rev{ \Pm \succeq \Om ,} 
\end{aligned}
\label{eq:convex_MIMO_diss}
\end{equation}
\rev{where $\gamma=\Pd/N_0$ is the \ac{SNR} when unit dissipated power is considered.}

\section{Dual problem}
\label{sec:dual}
One way of bounding the solution to~\eqref{eq:convex_MIMOr} is to construct a problem that will always have a solution greater than or equal to that of the initial problem. This problem is known as a dual to~\eqref{eq:convex_MIMOr}. The infimum of the dual problem provides an upper bound to the maximum of~\eqref{eq:convex_MIMOr} and they coincide when the duality gap is zero~\cite{Boyd+Vandenberghe2004}. \rev{The procedure for constructing the dual of~\eqref{eq:convex_MIMO_diss} is very similar to that of~\eqref{eq:convex_MIMOr}. For brevity we have chosen to include the derivation of the dual solution to~\eqref{eq:convex_MIMOr}, referring to App.~\ref{app:dissP_norm} for the dual of~\eqref{eq:convex_MIMO_diss}.  }

To construct a dual problem to~\eqref{eq:convex_MIMOr} we can combine the two constraints in~\eqref{eq:convex_MIMOr} into one, as a convex optimization problem with less constraints will always have a greater solution than the same problem with more constraints~\cite{Boyd+Vandenberghe2004}. 
A linear combination of the two constraints can be taken to restrict the dual problem
\begin{equation}
\Tr\left(\Rml\Pm + \nu\Rrm\Pm\right) = \nu + \rev{ \eta^{-1} -1},
\label{eq:linCombPar}
\end{equation}
where $\nu$ is a real scalar. Dividing the right-hand side to the left allows the introduction of a new matrix
\begin{equation}
\Rmn = \frac{\Rml + \nu\Rrm}{\nu + \rev{ \eta^{-1} -1}} .
\label{eq:Rmn}
\end{equation}
The dual to~\eqref{eq:convex_MIMOr} can now be written as
\begin{equation}
\begin{aligned}
	& \rev{\underset{\nu}{\mrm{min.}}\,\underset{\Pm}{\mrm{max.}}} && \log_2\det(\Id+\gamma\Sm\Pm\Sm^{\herm})\\
	& && \Tr(\Rmn\Pm) \rev{\leq} 1 \\
	& && \ \Pm \succeq \Om .
\end{aligned}	
\label{eq:convex_MIMO_dual}
\end{equation}
This problem is valid and convex for all values of $\nu$ for which $\Rmn$ is positive semi-definite\rev{, an interval calculated in App.~\ref{app:nu_interval}}. \rev{The solution to this problem will always be found at equality in its constraint due to the affine nature of the constraint. Therefore the optimization problem will be written with equality.}

To solve the dual problem, we rewrite it on such a form that it can be solved by water filling. This can be done by simplifying the condition restricting it. $\Rmn$ is positive semi-definite by construction and can therefore be Cholesky factorized as $\Rmn=\Bm^{\herm}\Bm$~\cite{Boyd+Vandenberghe2004}. By utilizing the cyclic invariance of the trace the condition can be rewritten as $\Tr(\Bm^{\herm}\Bm\Pm)=\Tr(\Bm\Pm\Bm^{\herm})$. The matrix $\Bm$ can be seen as a coordinate change for $\Pm$. This allows the introduction of a new variable $\Pmt=\Bm\Pm\Bm^{\herm}$ to write the optimization problem as,
\begin{equation}
\begin{aligned}
	&\rev{\underset{\nu}{\mrm{min.}}\,\underset{\Pmt}{\mrm{max.}}}&& \log_2\det(\Id+\gamma\Hmt\Pmt\Hmt^{\herm})\\
	& && \Tr(\Pmt) = 1 \\
	& && \ \Pmt \succeq \Om ,
\end{aligned}	
\label{eq:MIMOopt_newchan}
\end{equation}
where the new channel is $\Hmt=\Sm\Bm^{-1}$. The maximum of this problem can be found by water filling~\cite{Paulraj+etal2003}. To perform water filling it is a simple matter of following the same methodology outlined in~\cite{Paulraj+etal2003}, \ie find the \ac{SVD} of the channel matrix $\Hmt$ and iteratively fill the feeding vector $\Pmt$ such that the lowest loss channels are utilized the most. 
With the singular values of the channel matrix,~\eqref{eq:MIMOopt_newchan} can be written more simply as
\begin{equation}
\begin{aligned}
	& \rev{\underset{\nu}{\mrm{min.}}\,\underset{P_n}{\mrm{max.}}}&& \sum_{n=1}^N \log_2(1+\gamma\sigman^2 P_n)\\
	& && \sum_{n=1}^N P_n = 1 \\
	& && \ P_n \geq 0,
\end{aligned}	
\label{eq:waterfilling_opt}
\end{equation}
where $\sigman, \, n=1\dots N$ are the singular values of $\Hmt$\rev{, and $P_n$ is the power allocated to each mode associated with those singular values}. These singular values $\sigman$ can be expressed as, see App.~\ref{app:singval},
\begin{equation}
\rev{\sigman^2=\frac{(\nu +  \eta^{-1} -1)\rhon}{1+\nu\rhon}}
\rev{=\begin{cases}
(\eta^{-1}-1)\rhon, & \nu=0\\
1, & \nu\to\infty
\end{cases},}
	\label{eq:SVD_channel}
\end{equation}
where $\rhon$ are \rev{eigenvalues of modes} known as radiation modes~\cite{Schab2016} and are calculated through the \rev{generalized} eigenvalue problem,
\begin{equation}
\rev{\Rrm\Jm_n = \rhon \Rml\Jm_n ,}
\label{eq:eig_radmodes}
\end{equation}
\rev{where $\Jm_n$ are the radiation mode currents.}
These modes maximize the fraction between radiated power and power dissipated in ohmic losses, and have orthogonal currents and far-fields. It is evident from~\eqref{eq:SVD_channel} that these modes are a dominating factor in the singular values of the optimized channel. With this expression, water filling can be performed fast and efficiently with minimal numerical calculations. The minimization over $\nu$, which can be calculated using conventional minimizers such as \textit{fminbnd} in MATLAB, finally provides an upper bound to the initial problem~\eqref{eq:convex_MIMOr}
\rev{of the maximum spectral efficiency of MIMO antennas confined to the region $\reg$. The reformulation~\eqref{eq:waterfilling_opt} based on radiation modes~\eqref{eq:SVD_channel} also determines bounds on MIMO systems with $N$ ports by restricting the analysis to the $N$ strongest radiation modes. This restriction to $N$ modes can also be used to determine physical bound on the spectral efficiency for MIMO systems without channel state information utilizing equal power allocation~\cite{Paulraj+etal2003}. }

\rev{The singular values of the optimal channel when the problem is normalized to dissipated power, as in~\eqref{eq:convex_MIMO_diss}, can similarly be calculated as,}
\begin{equation}
   \rev{ \sigman^2 = \frac{(\nu+1-\eta)\rhon}{1+\nu(1+\rhon)}
   =\begin{cases}
(1-\eta)\rhon, & \nu=0\\
\displaystyle\frac{\rhon}{1+\rhon}, & \nu\to\infty
\end{cases},}
   \label{eq:SVD_channel_diss}
\end{equation}
\rev{see App.~\ref{app:dissP_norm}. }

\section{\rev{Radiation Modes}}
\label{sec:radmodes}
\rev{In Section~\ref{sec:dual} it is shown that the only part of the optimal spectral efficiency problem~\eqref{eq:convex_MIMOr} that is dependent on the geometry considered are the eigenvalues $\rhon$ of the radiation modes. This means that optimizing spectral efficiency with a radiation efficiency constraint has been reduced to solving the generalized eigenvalue problem~\eqref{eq:eig_radmodes}, followed by water filling over the singular values $\sigman$ given by~\eqref{eq:SVD_channel} or~\eqref{eq:SVD_channel_diss}. This means that it is possible to evaluate the quality of a geometry, in terms of spectral efficiency, by only studying the eigenvalues $\rhon$ of its radiation modes. This section details how $\rhon$ is connected to the spectral efficiency.
}   

\begin{figure}%
\includegraphics[width=\linewidth]{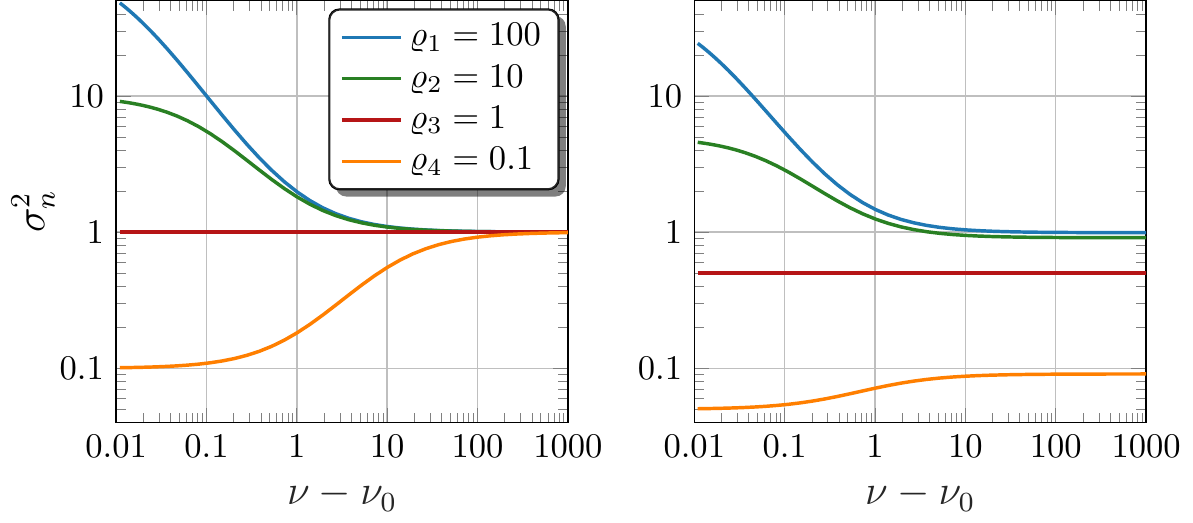}%
\caption{\rev{The singular values of the channel calculated for $\eta=0.5$ from~\eqref{eq:SVD_channel} for normalized radiated power on the left and from~\eqref{eq:SVD_channel_diss} for normalized dissipated power on the right. The $\nu$ scales have been adjusted by the start of the interval $\nu_0$, calculated in App.~\ref{app:nu_interval}, to include negative numbers.} }%
\label{fig:radmode_nu}%
\end{figure}
\rev{To understand how the spectral efficiency varies with $\rhon$ let us select some arbitrary values $\rhon=\{100,10,1,0.1\}$ and study the optimization problem. In Fig.~\ref{fig:radmode_nu} the singular values of the channel has been calculated by~\eqref{eq:SVD_channel} and~\eqref{eq:SVD_channel_diss} and plotted against the parameter $\nu$. Here we can see that a larger value of $\rhon$ corresponds to a stronger singular value of the channel $\sigman$, \ie less loss in the channel. For greater $\nu$ the singular value of the channel decreases for strong channels and increases for weak channels. When $\nu$ is varied the difference between the two different power normalizations become evident. For the radiated power normalization, all singular values converge to $1$ when $\nu$ increases~\eqref{eq:SVD_channel}. For the dissipated power normalization $\sigman$ stabilize with a difference proportional to $\rhon/(1+\rhon)$, see~\eqref{eq:SVD_channel_diss}. In Fig.~\ref{fig:radmode_nu} the eigenvalue $\varrho_3=1$ has a stable singular value for all $\nu$, this is due to the choice of $\eta$. The radiation efficiency restriction defines what is considered a good or a bad mode, separated by an eigenvalue for which the singular value is stable. This separating value can be found by studying the asymptotics of~\eqref{eq:SVD_channel} and~\eqref{eq:SVD_channel_diss} and is calculated as $\rhon=\eta/(1-\eta)$ for both normalized radiated and dissipated power. 
}

\begin{figure}%
\includegraphics[width=\linewidth]{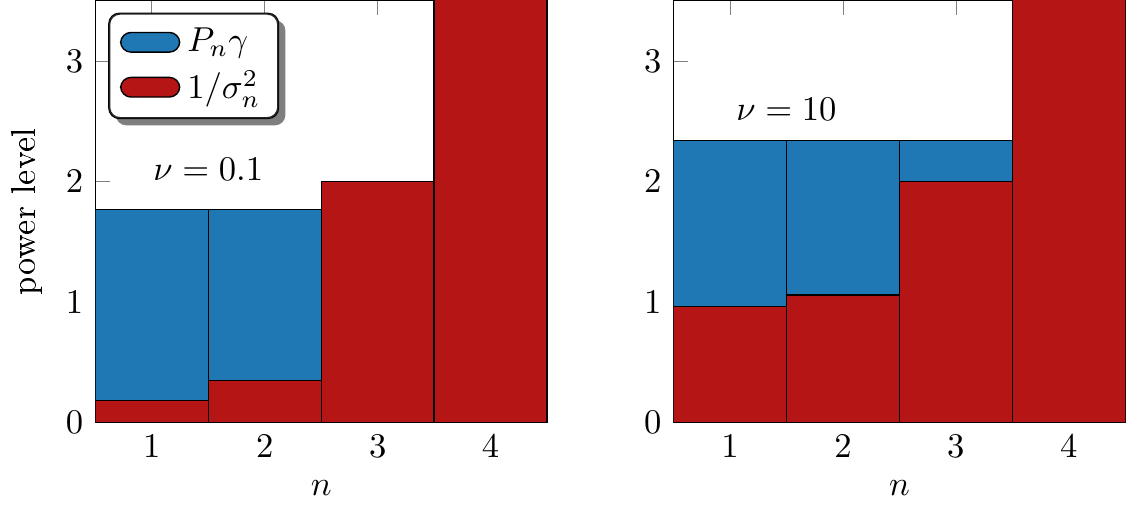}%
\caption{\rev{Water filling performed for the radiation mode eigenvalues $\rhon=\{100,10,1,0.1\}$ from Fig.~\ref{fig:radmode_nu}, while normalizing dissipated power. To the left $\nu=0.1$ and to the right $\nu=10$, with $\gamma=3$. The last bar extends far above the edge of the plot, due to the great amount of loss in the $\rho_4=0.1$ mode.}
}%
\label{fig:waterfill_nu}%
\end{figure}
\rev{To find the spectral efficiency, corresponding to different values of $\nu$, perform water filling on the singular values $\sigman$ found in Fig.~\ref{fig:radmode_nu}. This is illustrated in Fig.~\ref{fig:waterfill_nu} with $\nu=0.1$ and $\nu=10$, the \ac{SNR} set to $\gamma=3$, and dissipated power is normalized. The loss corresponding to each singular value is found by taking the inverse of its square, $1/\sigman^2$. The water filling algorithm then allocates power to the indices with least loss in the same way as pouring water into the graph. The resulting spectral efficiency is calculated from the first equation in~\eqref{eq:waterfilling_opt}. In  this example $\nu=0.1$ gives the spectral efficiency $C=3.1855$, $\nu=10$ gives $C=2.5829$, and the minimal value $C=2.5828$ is found at $\nu=\infty$. Here the smallest value is sought as it calculates the bound through~\eqref{eq:waterfilling_opt} to the initial problem of maximizing spectral efficiency in~\eqref{eq:convex_MIMO_diss}. As can be seen in Fig.~\ref{fig:radmode_nu}, for low $\nu$ the difference between the different singular values is increased leading to fewer channels being used. Whereas for high $\nu$ the difference between between the singular values decreases, leading to more channels being utilized, as in the right part of Fig.~\ref{fig:waterfill_nu}. In general, a low singular value will not be utilized unless the \ac{SNR} is very high, evident by $\rho_4=0.1$ producing a much greater loss than the other three in Fig.~\ref{fig:waterfill_nu}.}

\section{Results}
\label{sec:Results}
The solutions to the optimization problems~\eqref{eq:convex_MIMOr} \rev{and~\eqref{eq:convex_MIMO_diss}} \rev{provide upper bounds} on the spectral efficiency available for different structures. \rev{These bounds have} been verified by optimizing the same problems in CVX~\cite{Boyd+Vandenberghe2004,Grant+Boyd2011} for several of the considered cases confirming that the duality gap is zero for these cases. However, it is also interesting to investigate the radiation modes that contribute to the spectral efficiency in~\eqref{eq:SVD_channel}. To illustrate both of these results this section is divided into four sub-sections. First, the mode strength of different shapes is studied in Sec.~\ref{sec:Mode_strength}, then feeding a plate through sub-regions is investigated in Sec.~\ref{sec:Plate_sub-regions}. The optimal spectral efficiency of the same plate is discussed in Sec.~\ref{sec:optimal_cap}, and finally the mode availability of \rev{electrically} large shapes is shown in Sec.~\ref{sec:mode_availability}.

\subsection{Mode strength}
\label{sec:Mode_strength}
\begin{figure}%
\includegraphics[width=\columnwidth]{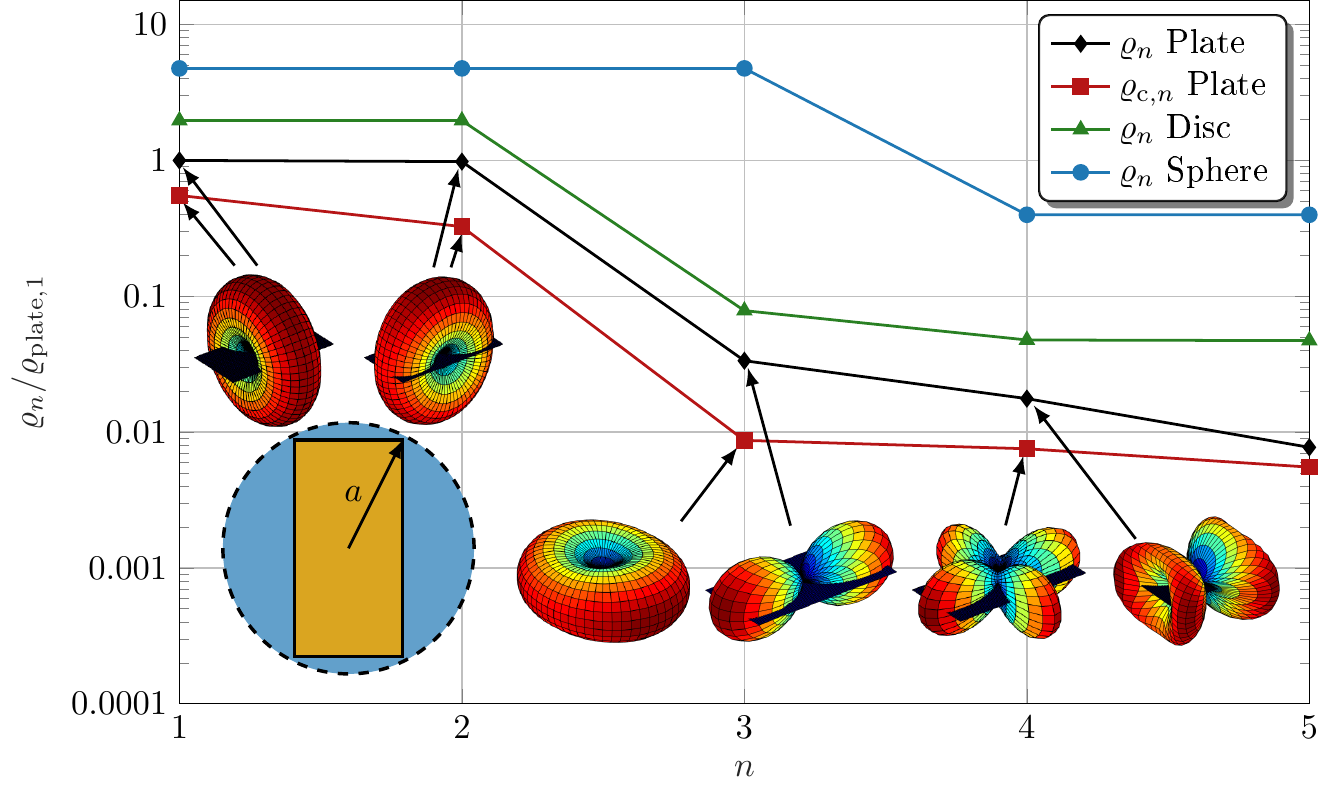}%
\caption{The \rev{relative} strength of the first $5$ radiation, $\varrho_n$, and characteristic, $\varrho_{\mrm{c},n}$, modes of a $ka = \rev{0.56}, \, \ell\times\ell/2$ plate, as well as the radiation modes of its circumscribing disc and sphere. All values have been normalized to \rev{the eigenvalue of} the first radiation mode of the plate \rev{, $\varrho_{\mrm{plate},1}\approx971$}. The corresponding far-fields for the radiation and characteristic modes of the plate can be seen as insets in the figure. The first two radiation and characteristic modes are visually identical and represented by the same patterns.}%
\label{fig:modeS_full}%
\end{figure}

In the derivation of the optimized channels singular values~\eqref{eq:SVD_channel}, the \rev{eigenvalues of the} radiation modes~\eqref{eq:eig_radmodes} are the only contributing factors that depend on the geometry of the structure. All other parameters are 
design specifications. The influence of antenna design and geometry on the problem is therefore fully described by these radiation modes. Their relative strength can \rev{serve as a measure of how many different modes are available for different shapes.}
\rev{This is the number of} orthogonal modes \rev{that} are available to provide diversity for those structures. In Fig.~\ref{fig:modeS_full} the relative strength of the radiation and \rev{loss-less} characteristic modes~\cite{Harrington1961,Chen+Wang2015} of a $\rev{ka=0.56}, \, \ell\times\ell/2$ plate, and the radiation modes of the circumscribing disc and sphere are shown normalized to the \rev{eigenvalue of the} first radiation mode of the plate. We see that the two first radiation modes have the same strength and the third and higher order modes are significantly weaker\rev{, and therefore require more input power to be utilized, or an increase in the plates size to be effective.} 
The characteristic mode strengths of the plate have been evaluated \rev{by putting their eigenvalues into the Rayleigh quotient related to the generalized eigenvalue problem~\eqref{eq:RadmodeEig}}, \ie
\begin{equation}
\rev{\varrho_{\mrm{c},n} = \frac{\Jm_{\mrm{c},n}^{\herm}\Rrm\Jm_{\mrm{c},n}}{\Jm_{\mrm{c},n}^{\herm}\Rml\Jm_{\mrm{c},n}} },
\label{eq:CM}
\end{equation}
where $\Jm_{\mrm{c},n}$ are the characteristic mode currents.
\rev{These are presented as a comparison to show how the characteristic modes perform at the task of maximizing spectral efficiency while maintaining high radiation efficiency.} We can see that the characteristic modes perform slightly worse than the radiation modes, and have different radiation patterns for higher order modes. The relative strength of radiation modes for the circumscribing disc and sphere, normalized to the \rev{eigenvalue of the} first radiation mode of the plate, have been included as a reference. We see that the disc has slightly higher mode strengths than the plate, whereas the sphere's modes are much stronger and grouped into a set of three for the first modes, rather than two.

\subsection{Plate sub-regions}
\label{sec:Plate_sub-regions}
\begin{figure}%
\includegraphics[width=\columnwidth]{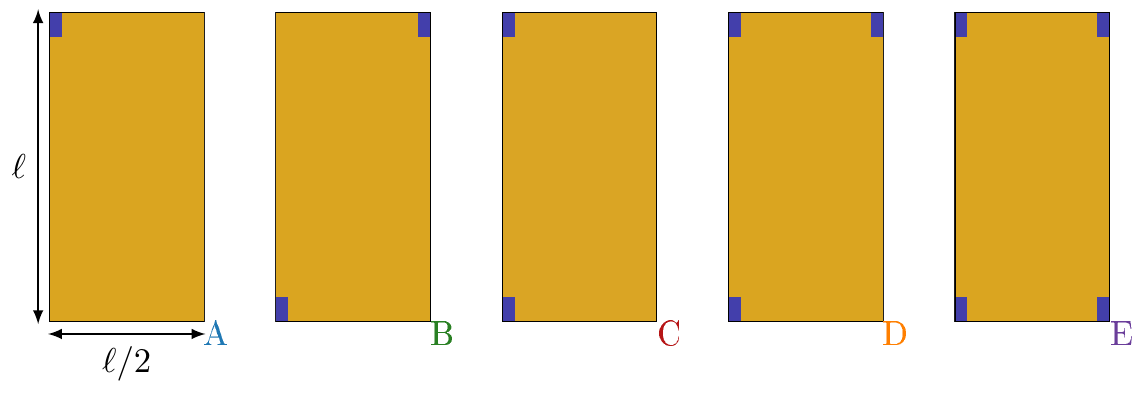}%
\caption{\rev{The sub-region cases studied in the examples. The small blue regions have the size $0.1\ell\times 0.05\ell$, each taking up $1\%$ of the total plate area, and are the regions in which the current is controlled during optimization.}}%
\label{fig:subreg}%
\end{figure}
\begin{figure}%
\includegraphics[width=\columnwidth]{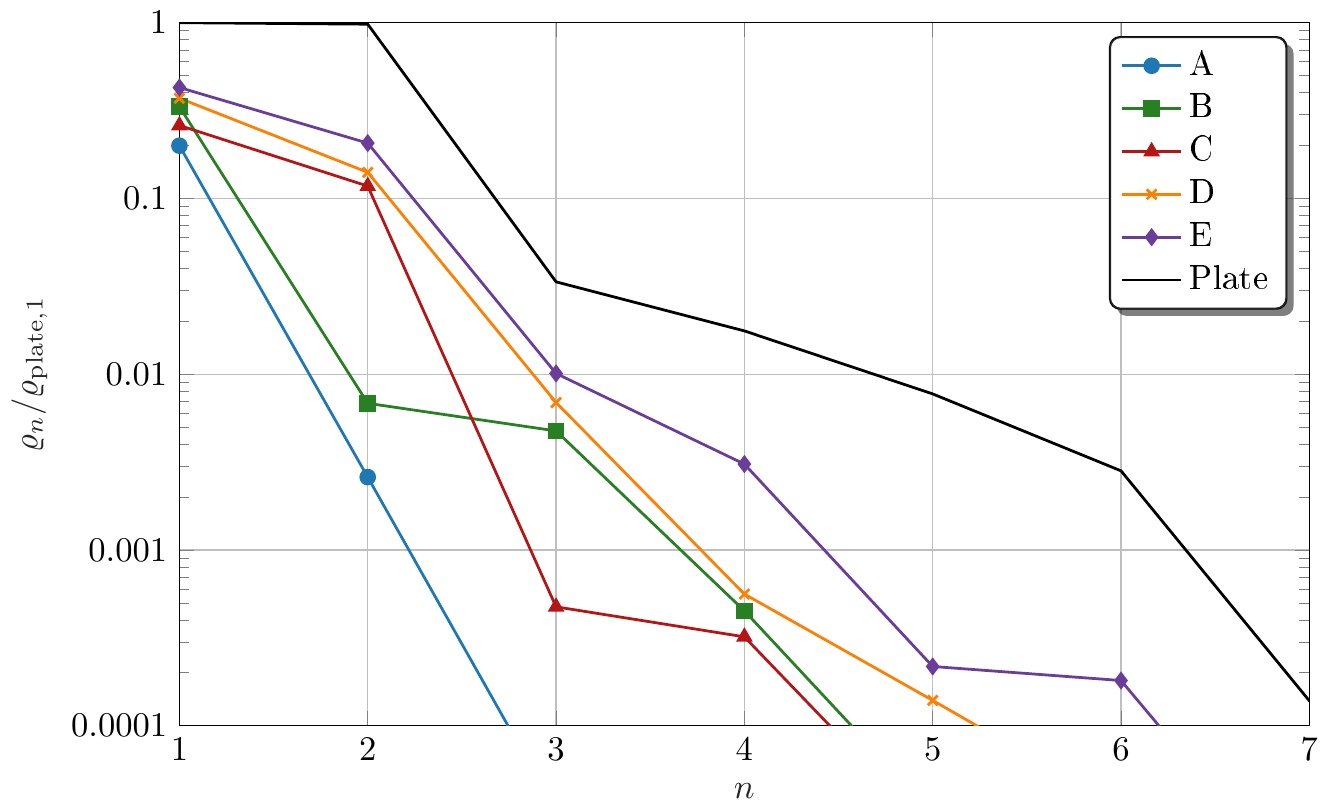}%
\caption{The \rev{relative} strength of each \rev{radiation} mode in Fig.~\ref{fig:modeS_full} when controlling only the current on sub regions \rev{shown in Fig.~\ref{fig:subreg}} of the $\rev{ka = 0.56, \, \ell\times\ell/2}$ plate with $R_{\mrm{s}}=0.01\,\Omega/\Box$\rev{, normalized to the eigenvalue of the first mode of the full plate}. 
The black curve without marks is the strength of each mode for the full plate normalized to the first mode.
}%
\label{fig:modeS_sub}%
\end{figure}

In wireless communication only a small part of the device is typically dedicated to antenna design. It is therefore interesting to see how well small sub-regions can excite the diversity available from the plate in Fig.~\ref{fig:modeS_full}. The optimization problem in Sec.~\ref{sec:dual} can be reformulated for a sub-region of a geometry with the rest of the volume acting as a ground plane, see App.~\ref{app:subregions}. In Fig.~\ref{fig:modeS_sub}, the sub-region problem has been solved for several different orientations of sub-regions\rev{, shown in Fig.~\ref{fig:subreg},} on the plate in Fig.~\ref{fig:modeS_full}, where each region covers $1\,\%$ of the plate's total area.
\rev{In this figure the eigenvalues have been normalized to the first eigenvalue of the full plates radiation mode. This means that the shape of Fig.~\ref{fig:subreg} is similar for different sized plates except for a change in relative magnitude. For reference the eigenvalue of the first radiation mode of the full plate is $\varrho_{\mrm{plate},1}\approx\{971,127,32\}$ for $ka=\{0.56,0.2,0.1\}$, or approximately $\varrho_{\mrm{plate},1}\approx k^2 A Z_0/(6\pi R_{\mrm{s}})$, where $Z_0$ is the free space impedance and $A=\ell^2/2$ the area. Multiplying these numbers to Fig.~\ref{fig:modeS_sub} gives an idea of what modes are effective for different sizes.}
Here, we can see how well different configurations of sub-regions are able to induce the diversity available in the plate when fed optimally. It is clear that a single sub-region, in case A, is only able to effectively induce the first radiation mode. However, two diagonally situated sub-regions, as in case B, are only marginally better at inducing the second mode. This is due to that the first two radiation and characteristic modes are induced diagonally across the plate~\cite{Li+etal2014}. Therefore, the diagonally opposite regions do not effectively induce the second diagonal mode. However, if the two regions are placed on the same side of the plate, as in case C, the second order mode is induced effectively. The radiation mode strengths are very similar if these two regions are placed on the long side of the plate. When going to higher order modes, this configuration is no longer as effective, here, the diagonal regions in case B dominate. If three or four sub-regions are utilized, as in cases D and E, we can get both of these properties. However, adding the fourth sub-region only marginally increases the strength of the three first radiation modes. 

The values in Fig.~\ref{fig:modeS_sub} have a negligible dependence on the surface resistance $\Rs$\rev{, because they are normalized to the eigenvalue of the first radiation mode}. However, the calculation to produce the sub-region problem is dependent on the full impedance matrix $\Zm$ and thus the surface resistance, see App.~\ref{app:subregions}. This marginally changes the relation between the mode strength. However, the surface resistance plays an important role in which modes are utilized when feeding the structure for optimal spectral efficiency. 
When the surface resistance is increased the loss in the higher order modes is increased, \rev{pushing higher order modes out of availability. Therefore decreasing the number of modes that can be utilized.} 


\subsection{Optimal spectral efficiency}
\label{sec:optimal_cap}
\begin{figure}%
\includegraphics[width=\columnwidth]{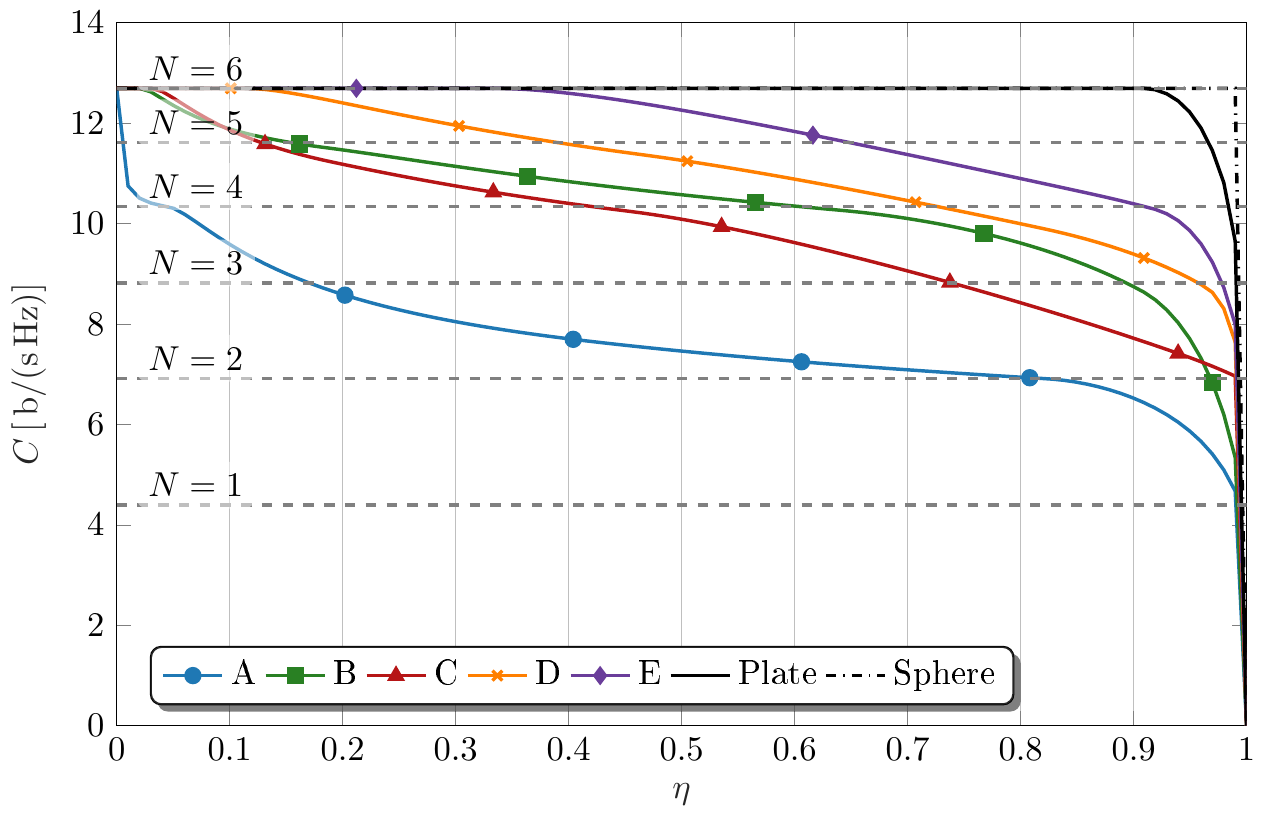}%
\caption{\rev{The optimal spectral efficiency for normalized radiated power~\eqref{eq:SVD_channel} for a $ka=0.56, \, \ell\times\ell/2$ plate, its circumscribing sphere, and the sub-region cases in Fig.~\ref{fig:subreg} for different required radiation efficiencies. The \ac{SNR} is fixed to $\gamma=20$. The horizontal lines show the spectral efficiency of equal power allocation to a certain number, $N=\{1,2,3,4,5,6\}$, of ideal channels. The physical objects have been restricted to use at most 6 different modes.}
}%
\label{fig:CapvsEff_ka=0p56_rad_ch_res}%
\end{figure}
\rev{In Fig.~\ref{fig:CapvsEff_ka=0p56_rad_ch_res}, the optimization problem~\eqref{eq:convex_MIMOr} restricted to MIMO systems with $N=6$ ports has been solved normalized to radiated power for different required radiation efficiencies.  This is compared in the figure to equal power allocation to a set number of ideal, loss-less channels. We can see that $6$ such channels perfectly bounds all considered geometries when they are at most using $6$ radiation modes. With the different curves crossing the number of ideal channels when they start to utilize that number of modes. The number of used modes can be compared to Fig.~\ref{fig:modeS_sub}, where we can predict how many effective modes the different sub-region orientations have. Multiply the numbers in Fig.~\ref{fig:modeS_sub} with $\varrho_{\mrm{plate,}1}\approx971$ and judge an effective mode as one with an eigenvalue above $\eta/(1-\eta)$. We see that case A has around 2 effective modes, which is the region in which it is situated in Fig.~\ref{fig:CapvsEff_ka=0p56_rad_ch_res}. Similarly we see that case B has 3 effective modes and one slightly lower mode, whereas C has 2 effective modes and two modes slightly below the effective line, this is mirrored in the optimal spectral efficiency where B slightly outperforms C. In this way we can predict the results of Fig.~\ref{fig:CapvsEff_ka=0p56_rad_ch_res} by studying the eigenvalues of the radiation modes in Fig.~\ref{fig:modeS_sub}.
The real world implication of this result is that to feed a certain degree of \ac{MIMO} system that number of radiation modes must be effective for the geometry. Take for example a two port antenna design aiming to effectively induce two orthogonal channels across its shape for communication. The viability of different designs for this purpose can be evaluated by studying the values of the eigenvalues for the two first radiation modes. These curves look slightly different if dissipated power is normalized. In general each sub-region configuration is stable over all required radiation efficiencies, dropping off once the solution is no longer realizable, see~\cite{Ehrenborg+Gustafsson2018} for an example. The reason for this low dependence on $\eta$ can be found in the asymptotic of~\eqref{eq:SVD_channel_diss}, where the singular values of the channel are independent of $\eta$ for large $\nu$.}

\begin{figure}
    \centering
    \includegraphics[width=\columnwidth]{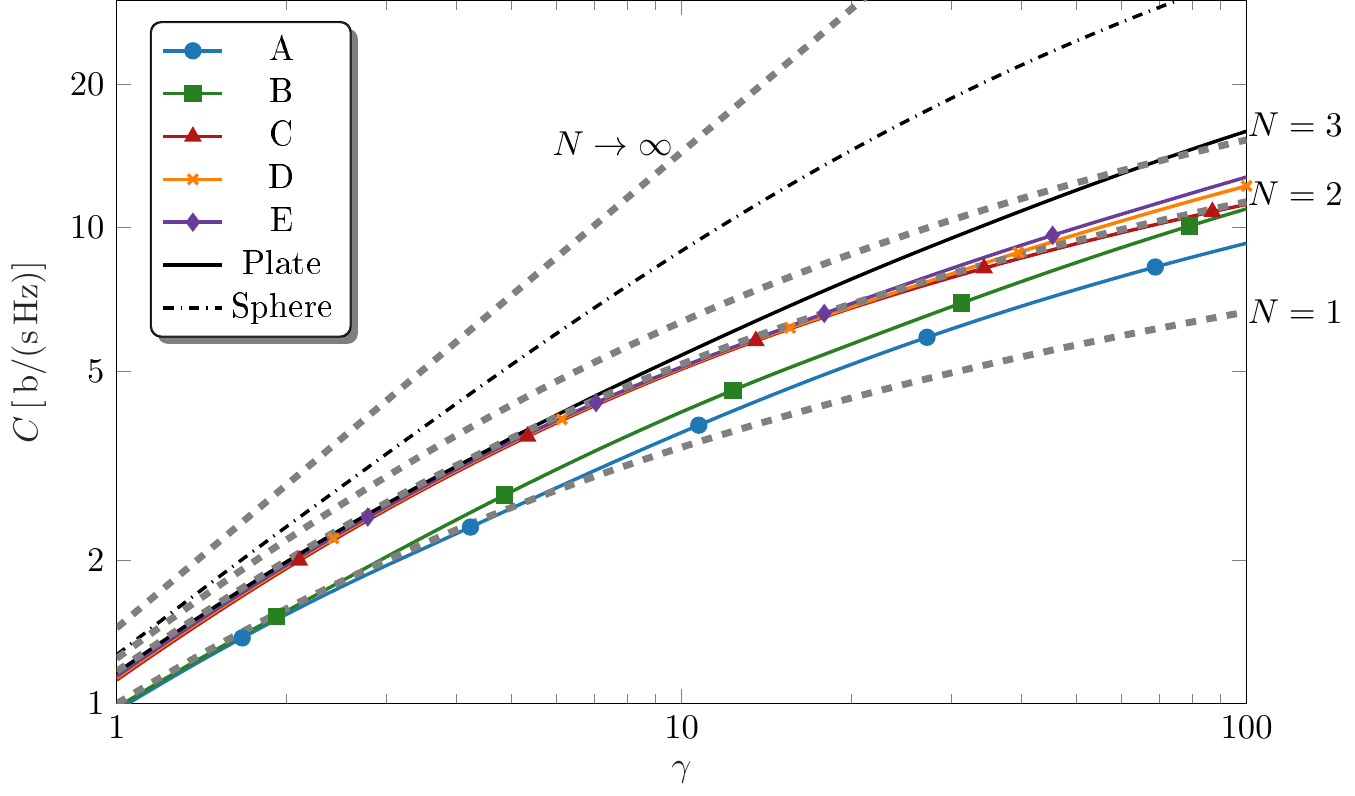}
    \caption{\rev{The optimal spectral efficiency for normalized dissipated power for a $ka=0.2, \, \ell\times\ell/2$ plate, its circumscribing sphere, and the sub-region cases in Fig.~\ref{fig:subreg} for different \ac{SNR}. The radiation efficiency is chosen as $\eta=0.5$. The dashed gray lines indicate  ideal, loss-less, equal power allocation channels with $N=\{1,2,3,\infty\}$.}}
    \label{fig:CapvsSNR_ka=0p2_diss}
\end{figure}
\rev{For the dissipated power normalization it is interesting to study the dependence on \ac{SNR} rather than efficiency. In Fig.~\ref{fig:CapvsSNR_ka=0p2_diss} the problem~\eqref{eq:convex_MIMO_diss} has been solved over a range of $\gamma$, for the plate in Fig.~\ref{fig:subreg}. It can be seen that for \ac{SNR} $\gamma<10$ the two cases A and B induce one mode whereas the other sub-region orientations, as well as the full plate, induce two. This result can be predicted by studying Fig.~\ref{fig:modeS_sub} where cases A and B have a much weaker second mode than the others. However, when we increase \ac{SNR} cases A and B start to approach the performance of two ideal channels.}

\begin{figure}%
\includegraphics[width=\columnwidth]{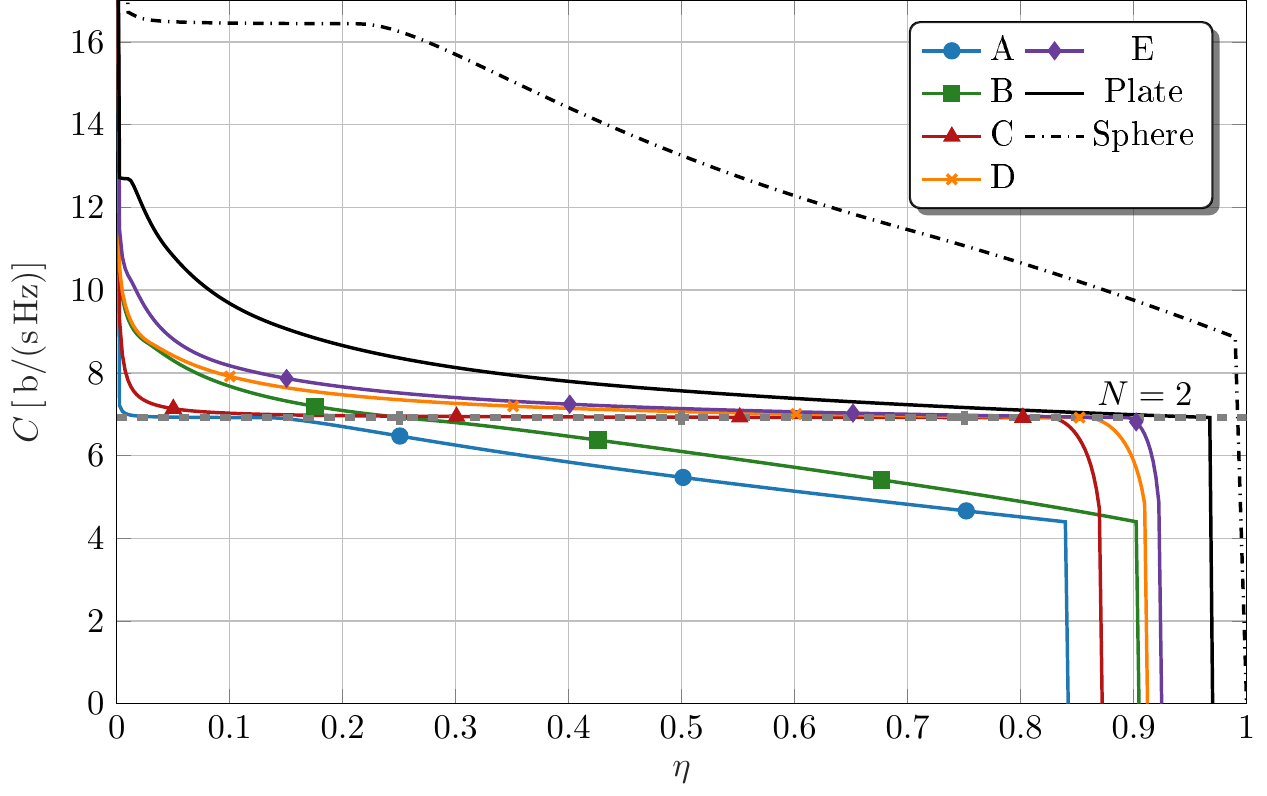}%
\caption{The optimal spectral efficiency with normalized radiated power for a $\rev{ka=0.1}, \, \ell\times\ell/2$ plate, its circumscribing sphere, and the sub-region cases in Fig.~\ref{fig:subreg} for different required radiation efficiencies. \rev{The \ac{SNR} is fixed to $\gamma=20$.}}%
\label{fig:CapvsEff_ka=0p1_rad}%
\end{figure}

\rev{The different sub-regions introduced in Fig.~\ref{fig:subreg} show some separation from the spectral efficiency that can be induced by the full plate in Fig.~\ref{fig:CapvsEff_ka=0p56_rad_ch_res}. However, when we consider a smaller plate of size $ka=0.1$, in Fig.~\ref{fig:CapvsEff_ka=0p1_rad}, we see that it is possible to induce almost as much spectral efficiency as the full plate when only optimizing the currents inside the small sub-regions. } %
\rev{This is true for all sub-region orientations except for cases A and B, which drop off faster than the others. We see that all other geometries produce as much spectral efficiency as $2$ ideal, loss-less, equal power allocation channels, indicating that they utilize two different modes.}
The cut-offs correspond to radiation efficiencies where a solution is no longer feasible. From this it is possible to infer that two or three cleverly placed regions is enough to induce all available spectral efficiency for a $\rev{ka=0.1}$ plate. Interestingly case C (two regions on the same side of the plate) has an earlier cut-off than case B (two diagonally opposed regions) even though case C outperforms B for lower radiation efficiency requirements. This most likely corresponds to the slightly higher mode strength of the first mode for case B seen in Fig.~\ref{fig:modeS_sub}. For this plate the optimal spectral efficiency of the sphere illustrates that the plate bound is much tighter for planar structures than that of the circumscribing sphere.

\subsection{Mode availability}
\label{sec:mode_availability}
The examples in Sec.~\ref{sec:Mode_strength},~\ref{sec:Plate_sub-regions},~\ref{sec:optimal_cap} have concerned \rev{electrically} small structures. The smaller sizes accentuate the availability of the lower order modes. When the size of the structure starts to grow in terms of wavelength a plethora of modes become effective, \ie have high associated eigenvalues. Through solving the eigenvalue problem~\eqref{eq:eig_radmodes} the number of viable modes, $\varrho_{\mrm{eff}}$, \rev{are those with an eigenvalue greater than $\eta/(1-\eta)$, see Sec.~\ref{sec:radmodes}.} This is seen as when the excitation of the mode does not accrue more losses. Since~\eqref{eq:eig_radmodes} depends on the geometry of the structure it is possible to analyze different shapes and study how many modes they have available. However, it is intuitive that shapes with greater surface area will induce more modes. To understand if these shapes are actually inducing these extra modes efficiently per area the number of modes can be normalized to the natural number of degrees of freedom for that area~\cite{Migliore2006a,Migliore2008,Hu+etal2017}. That number is defined as
\begin{equation}
2L(L+2) \approx 2ka(ka+2) \to 2(ka)^2 \, , \, ka \to \infty ,
\end{equation}
for a sphere~\cite{Harrington1961}. This term can be rewritten in terms of the area of the sphere $A$,
\begin{equation}
2(ka)^2 = \frac{k^2A}{2\pi}= \frac{2\pi A}{\lambda^2} .
\label{eq:degrees_freedom}
\end{equation}
\begin{figure}%
\includegraphics[width=\columnwidth]{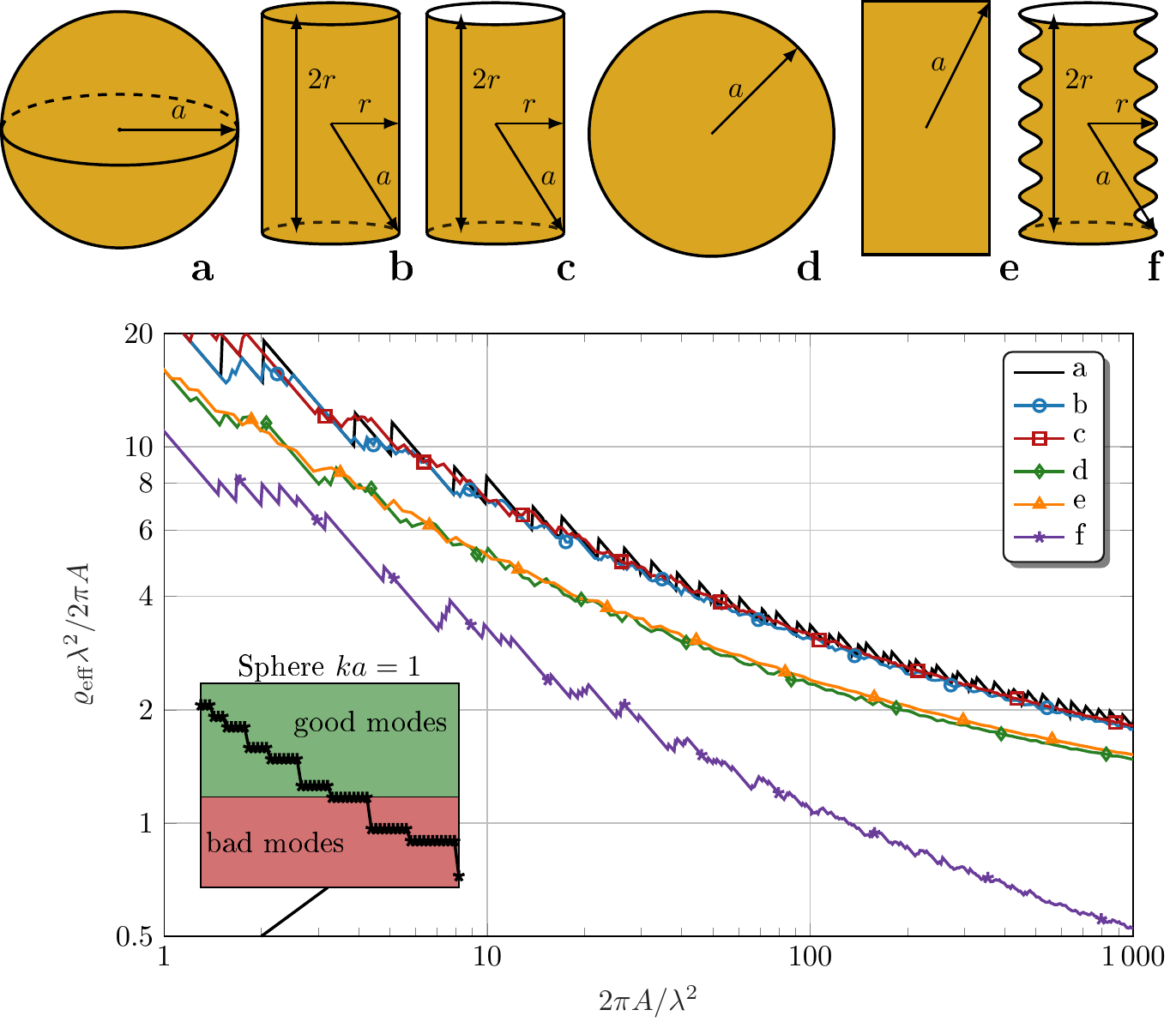}%
\caption{The number of viable modes, $\varrho_{\mrm{eff}}$, \rev{for $\eta=0.5$,} normalized to their natural number of degrees of freedom for different shapes. This is plotted against the surface area normalized to wavelength, see~\eqref{eq:degrees_freedom}. The surface resistance of the shapes is $\Rs=0.01\,\Omega/\Box$. \rev{The inset shows which of the radiation modes of a $ka=1$ sphere are considered effective.}}%
\label{fig:ModevsKa}%
\end{figure}
For the analysis of different shapes presented in Fig.~\ref{fig:ModevsKa} the degrees of freedom for each shape has been calculated by~\eqref{eq:degrees_freedom} using the surface area of each geometry\rev{, and $\eta=0.5$}. We can see that the curves are divided into two distinct groups, the three dimensional shapes \textbf{a}, \textbf{b}, and \textbf{c}, and the two dimensional \textbf{d} and \textbf{e}. This metric shows that a shape such as a cylinder induces modes almost as efficiently per area as a sphere. By arbitrarily permuting the surface of a shape to increase its area, such as for the cylinder \textbf{f}, we can see that the efficiency does not increase with increased area. In fact the efficiency of this shape is much lower. It is reasonable to conclude that convex shapes with maximum area, such as the sphere, will be able to induce the greatest number of modes efficiently. The jagged jumps in the curves are due to the fact that the \rev{eigenvalues of}
modes \rev{are} not continuously distributed. \rev{This is illustrated for a sphere in the inset in Fig.~\ref{fig:ModevsKa}, where we can see that there is a significant jump between the eigenvalues for different groups of modes.}
The consequences of this is that the number of efficient modes is monotonically increasing with frequency, but when it is normalized with the wavelength, as in Fig.~\ref{fig:ModevsKa}, discrete jumps will occur when new groups of modes become available. For closed shapes, such as the sphere, the internal resonances\rev{, occurring in the \ac{MoM} evaluation of the modes,} also reduce the number of efficient radiation modes for certain sizes.

\section{Conclusions}
\label{sec:conclusions}
In this paper a \rev{computationally efficient}
expression based on water-filling and radiation modes for calculating an upper bound on the optimal spectral efficiency for an arbitrary shape, constrained by the radiation efficiency, was presented.  It was illustrated that the \rev{radiation modes can serve as a metric for studying spectral efficiency.} 
The \rev{eigenvalues, relative} strength, and availability of the radiation modes were studied for several shapes including a plate, disc, cylinder, and sphere. Through this analysis it was shown that it is possible to excite currents producing optimal spectral efficiency for a plate using only a few small sub-regions. The number of effective modes for electrically large shapes was investigated, and it was illustrated that cylindrical shapes produce roughly as many effective modes per surface area as a sphere.

\rev{The method presented here can also be used to study capacity optimization when there is no channel state information. For such a situation equal power allocation would be used for a number of radiation modes over structure. Optimizing this number would yield the optimal solution. There is also potential to extend this method to treat additional constraints in a computationally efficient manner, such as self resonance, or Q-factor constraints. It remains as an interesting future prospect to include random propagation channels and other realizations into the optimization.}

\begin{appendices}
\section{The $\Sm$ matrix}
\label{app:Smatrix}
The elements of the loss less  impedance matrix are calculated by the integral
\begin{equation}
Z_{pq} = \mrm{j} k Z_{0} \int_{\mrm{\varOmega}} \int_{\mrm{\varOmega}} \basfcn_p \left(\V{r}_1\right) \cdot \mat{G} \left(\V{r}_1,\V{r}_2\right) \cdot \basfcn_q \left(\V{r}_2\right) \diff A_1 \diff A_2 ,
\label{eq:momZ}
\end{equation}
where $Z_{0}$ is the free wave impedance, $\mrm{\varOmega}$ is the source region, $\basfcn_p$ and $\basfcn_q$ are the basis and test functions on the antenna, and $\diff A$ denotes the area element integrated over~\cite{Chew+etal2008}. 
The Greens dyadic inside of this expression can be written as a product between out-going and regular spherical vector waves. The radiation matrix can be found by taking $\Rrm=\Re(\Zm)$. By taking the real value of~\eqref{eq:momZ} both of the spherical vector waves become regular and it is possible to split the double integral into two identical integrals,
\begin{equation}
S_{\alpha p} = k\sqrt{Z_0}\int_{\mrm{\varOmega}} \basfcn_p \left(\V{r}\right)\,\cdot\,\vec{u}_{\alpha}^{(1)}(k\V{r})\diff A ,
\end{equation}
where $\vec{u}_{\alpha}^{(1)}$ are the regular spherical vector waves~\cite{Tayli+etal2018}. Those integrals produce a matrix denoted as $\Sm$ which is the matrix connecting the basis functions in $\mrm{\varOmega}$ to the spherical modes in the far-field. The radiation matrix can thus be decomposed as $\Rrm=\Sm^{\herm}\Sm$. 

\section{SVD of $\Sm\Bm^{-1}$}
\label{app:singval}
The singular values of a matrix can be calculated by taking the positive square root of the eigenvalues of the matrix times itself, $(\eig(\Hmt\Hmt^{\herm}))^{1/2}$. By expanding the channel matrix to its component matrices,
\begin{equation}
\eig(\Sm\Bm^{-1}\Bm^{-\herm}\Sm^{\herm}) = \eig(\Sm\Rmn^{-1}\Sm^{\herm}) ,
\label{eq:channel_eig}
\end{equation}
it can be seen that \rev{the evaluation of} $\Rmn^{-1}=(\nu + \rev{ \eta^{-1} -1})(\Rml + \nu\Rrm)^{-1}$ needs to be 
independent of $\nu$. Due to their good properties, it is possible to decompose both $\Rml$ and $\Rrm$ into more manageable matrices. Let's start with $\Rrm$, this matrix is the real valued part of the \ac{MoM} impedance matrix $\Zm$ in~\eqref{eq:momZ}. It is possible to decompose $\Rrm$ as a multiplication between two instances of the $\Sm$ matrix, $\Rrm = \Sm^{\herm}\Sm$, see Appendix~\ref{app:Smatrix}. The loss matrix can be decomposed using a Cholesky factorization $\Rml=\Upsilonm^{\herm}\Upsilonm$. \rev{These matrices are numerically efficient to handle as $\Sm$ is a low rank matrix and $\Upsilonm$ is a sparse matrix.}
\rev{With these decompositions the matrix inside the eigenvalue problem~\eqref{eq:channel_eig} can now be written as,}
\begin{multline}
     \rev{\Sm\Rm_{\nu}^{-1}\Sm^{\herm}} \\
  \rev{=(\nu + \rev{ \eta^{-1} -1})\Sm\Upsilonm^{-1}(\Id+\nu\Upsilonm^{-\herm}\Sm^{\herm}\Sm\Upsilonm^{-1})^{-1}\Upsilonm^{-\herm}\Sm^{\herm}.}
  \label{eq:square_channel}
\end{multline}
\rev{Take an \ac{SVD} of the two decomposition matrices, $\Sm\Upsilonm^{-1}=\Um\Sigmam\Vm^{\herm}$, and simplify~\eqref{eq:square_channel},}
\begin{equation}
    \rev{\Sm\Rm_{\nu}^{-1}\Sm^{\herm}} 
  \rev{=(\nu + \rev{ \eta^{-1} -1})\Um\Sigmam(\rev{\Id}+\nu\Sigmam^{2})^{-1}\Sigmam\Um^{\herm}.}
\end{equation}


Putting this back into the eigenvalue problem~\eqref{eq:channel_eig} \rev{an efficient solution with no numerical operations dependent on $\nu$ is found,} 
\begin{equation}
  \rev{\eig(\Sm\Rm_{\nu}^{-1}\Sm^{\herm})
  =\frac{(\nu +  \eta^{-1} -1)\rhon}{1+\nu\rhon}, }
\label{App_eq:SVD_channel}
\end{equation}
where $\rhon$ are the eigenvalues of the generalized eigenvalue problem,
\begin{equation}
\Rrm\Jm_n = \rhon\Rml\Jm_n.
\end{equation}
The solutions to this eigenvalue problem are known as radiation modes~\cite{Schab2016}. 
These are related to the singular values of $\Sm\Upsilonm^{-1}$ as,
\begin{multline}
\svd(\Sm\Upsilonm^{-1}) = (\eig(\Sm\Rml^{-1}\Sm^{\herm}))^{1/2}\\
 = (\eig(\Rrm,\Rml))^{1/2} = \rhon^{1/2}.
\label{eq:RadmodeEig}
\end{multline}
It is evident that one of the main contributions to the singular values of the optimal channel are \rev{the eigenvalues of} these radiation modes. In fact, the only other part of~\eqref{App_eq:SVD_channel} is a requirement on \rev{the radiation efficiency $\eta$}. This is a parameter which the designer does not, in general, have control over. However, the \rev{eigenvalues}
of the radiation modes, $\rhon$, depend on the geometry of the structure, which is controllable.

\section{$\nu$ interval}
\label{app:nu_interval}
The linear combination of the two conditions in~\eqref{eq:convex_MIMOr} are valid for values of $\nu$ that ensure that the resulting matrix $\Rmn$ is positive semidefinite. Since both of its constituent matrices are positive definite this implies that $\nu>\nu_{0}$\rev{, where $\nu_0$ is the lower limit at which $\Rmn$ is positive semidefinite.} The lower \rev{limit} can be established by studying the eigenvalue problem in~\eqref{eq:SVD_channel},
\begin{equation}
\frac{(\nu + \rev{ \eta^{-1} -1})\rhon}{\rev{1}+\nu\rhon}\geq 0 ,
\end{equation}
\rev{where} $\rhon$ is always positive since it is a generalized eigenvalue of two positive semi-definite matrices, \rev{and the efficiency $\eta$ is a constant between $[0,1]$. Therefore we gain two conditions that must be fulfilled for $\Rmn$ to be positive semidefinite.,} 
\begin{equation}
    \rev{\nu\geq 1-\eta^{-1}
    \quad\text{and }
    \nu\geq -\frac{1}{\rhon}} ,
\end{equation}
\rev{from the nominator and denominator, respectively.}
\rev{
The greatest of the two will provide the limit, \ie $\nu_0=\max{\{-\rhon^{-1},1-\eta^{-1}\}}$.}

\section{Dissipated power normalization}
\label{app:dissP_norm}
\rev{When formulating the dual problem and solution to~\eqref{eq:convex_MIMO_diss} we follow the same steps detailed in Section~\ref{sec:dual} and Appendix~\ref{app:singval}, with some differences in the matrix $\Rmn$.  } 
\rev{When we combine the two matrices in~\eqref{eq:convex_MIMO_diss} we get, }
\begin{equation}
\Tr\left(\left(\Rml+\nu(\Rrm+\Rml)\right)\Pm\right) = \nu + \rev{1-\eta},
\end{equation}
which can be simplified by the introduction of a new matrix, as in~\eqref{eq:Rmn},
\begin{equation}
\Rmn = \frac{1}{\nu + \rev{1-\eta}}\left(\Rml+\nu(\Rrm+\Rml)\right) .
\end{equation}
\rev{If this matrix is used in the derivation detailed in Appendix~\ref{app:singval}, the eigenvalues are calculated through the expression,}
\begin{equation}
    \rev{\eig(\Sm\Rm_{\nu}^{-1}\Sm^{\herm}) = \frac{(\nu+1-\eta)\rhon}{1+\nu(1+\rhon)} .} 
\end{equation}
The lower limit on $\nu$ is thus calculated \rev{by the expressions,}
\begin{equation}
\rev{\nu \geq \eta - 1
\quad\text{and }
\nu \geq -\frac{1}{1+\rhon}}
\end{equation}
\rev{from the nominator and denominator, respectively.}
\rev{
Taking the maximum of these expressions as the limit, $\nu_0=\max{\{-(1+\rhon)^{-1},\eta - 1\}}$.} 


\section{Sub-regions}
\label{app:subregions}
In order to simulate embedded antennas the antenna problem must be reformulated in the currents that are controlled~\cite{Gustafsson+Nordebo2013}. Consider the \ac{MoM} matrix formulation $\Zm \Jm = \Vm$, divide it into sub-matrices related to the controlled currents, denoted to subscript a, and induced currents, denoted by subscript g,
\begin{equation}
\begin{pmatrix}
	\Zm_{\mrm{aa}} & \Zm_{\mrm{ag}}\\
	\Zm_{\mrm{ga}} & \Zm_{\mrm{gg}}
\end{pmatrix}
\begin{pmatrix}
	\Jm_{\mrm{a}} \\
	\Jm_{\mrm{g}}
\end{pmatrix}
=
\begin{pmatrix}
	\Vm_{\mrm{a}} \\
	\Om
\end{pmatrix} ,
\label{eq:SubZVI}
\end{equation}
where $\Zm_{\mrm{ag}}$ connects the controlled region to the induced region, and $\Zm_{\mrm{ga}}$ connects the induced region to the controlled region. The right hand side is only non-zero for the controlled region. The second equation is used to express the induced currents in terms of the controlled ones,
\begin{equation}
\Jm_{\mrm{g}} = -\Zm_{\mrm{gg}}^{-1}\Zm_{\mrm{ga}}\Jm_{\mrm{a}} = \Zm_{\mrm{t}}\Jm_{\mrm{a}} .
\label{eq:passiveCurrents} 
\end{equation}
The \ac{MoM} matrices of the problem can now be reformulated into forms which only act on controlled currents. Take the Gram matrix $\Psim$ as an example,
\begin{multline}
\Jm_{\mrm{a}}^{\rev{\herm}}\Psim\Jm_{\mrm{a}} = \Jm_{\mrm{a}}^{\herm}\Psim_{\mrm{aa}}\Jm_{\mrm{a}} + \Jm_{\mrm{a}}^{\herm}\Psim_{\mrm{ag}}\Jm_{\mrm{g}} + \Jm_{\mrm{g}}^{\herm}\Psim_{\mrm{ga}}\Jm_{\mrm{a}} + \Jm_{\mrm{g}}^{\herm}\Psim_{\mrm{gg}}\Jm_{\mrm{g}} \\
= \Jm_{\mrm{a}}^{\herm}(\Psim_{\mrm{aa}} +
\rev{\Psim_{\mrm{ag}}\Zm_{\mrm{t}}+\Zm_{\mrm{t}}^{\herm}\Psim_{\mrm{ga}}} +
 \Zm_{\mrm{t}}^{\herm}\Psim_{\mrm{gg}}\Zm_{\mrm{t}})\Jm_{\mrm{a}} \\
= \Jm_{\mrm{a}}^{\herm}\Psim_{\mrm{p}}\Jm_{\mrm{a}} .
\end{multline}
Similarly the $\Sm$ matrix connecting the currents to the spherical waves in the far-field can be rewritten as,
\begin{equation}
\Sm^{\herm}\Jm = \Sm^{\herm}_{\mrm{a}}\Jm_{\mrm{a}} + \Sm^{\herm}_{\mrm{g}}\Jm_{\mrm{g}} = \Sm^{\herm}_{\mrm{a}}\Jm_{\mrm{a}} + \Sm^{\herm}_{\mrm{g}}\Zm_{\mrm{t}}\Jm_{\mrm{a}} = \Sm^{\herm}_{\mrm{p}}\Jm_{\mrm{a}} .
\end{equation}

\end{appendices}


\begin{IEEEbiography}[{\includegraphics[width=1in,height=1.25in,clip,keepaspectratio]{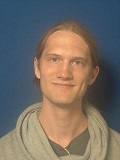}}]{Casimir Ehrenborg (S'15)} 
received his M.Sc. degree in engineering physics  from Lund University, Sweden, in 2014. He is currently a Ph.D. student in the Electromagnetic Theory Group, Department of Electrical and Information Technology, Lund University. In 2015, he  participated in and won the IEEE Antennas and  Propagation Society Student Design Contest for his body area network antenna design. In 2019 he received the IEEE Uslenghi letters prize paper award. His research interests include small antennas, stored energy, phase and radiation centers, as well as physical bounds.
\end{IEEEbiography}

\begin{IEEEbiography}[{\includegraphics[width=1in,height=1.25in,clip,keepaspectratio]{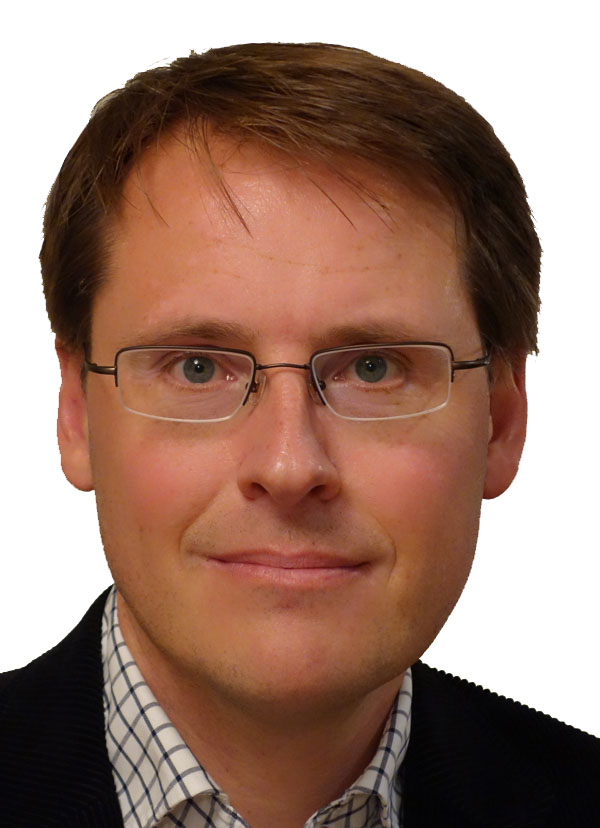}}]{Mats Gustafsson}
received the M.Sc. degree in Engineering Physics 1994, the Ph.D. degree in Electromagnetic
Theory 2000, was appointed Docent 2005, and Professor of Electromagnetic Theory 2011, all
from Lund University, Sweden.

He co-founded the company Phase holographic imaging AB in 2004. His research interests are in scattering and antenna theory and inverse scattering and imaging. He has written over 90 peer reviewed journal papers and over 100 conference papers. Prof. Gustafsson received the IEEE Schelkunoff Transactions Prize Paper Award 2010 and Best Paper Awards at EuCAP 2007 and 2013. He served as an IEEE AP-S Distinguished Lecturer for 2013-15.
\end{IEEEbiography}
%

\end{document}